\documentstyle[epsfig]{mn}
\begin{document}
\def\Msun{M_{\odot}}
\def\simlt{\mathrel{\rlap{\lower 3pt\hbox{$\sim$}}\raise 2.0pt\hbox{$<$}}}
\def\simgt{\mathrel{\rlap{\lower 3pt\hbox{$\sim$}} \raise 2.0pt\hbox{$>$}}}

\title[Cosmic Backgrounds from Miniquasars]{Cosmic Backgrounds from Miniquasars}
\author[Salvaterra, Haardt \& Ferrara]
{Ruben Salvaterra$^{1}$, Francesco Haardt$^{1}$  \& Andrea Ferrara$^{2}$  \\
$^1$Dipartimento di Fisica e Matematica, Universit\'a dell'Insubria, Via Valleggio 11, 22100 Como, Italy\\
$^2$SISSA/International School for Advanced Studies, Via Beirut 4, 34100 Trieste, Italy}

\maketitle \vspace {7cm}

\begin{abstract}
A large population of Intermediate Mass Black Holes (IMBHs) might be produced at early cosmic
times as a left over of the evolution of the very massive first stars. Accretion onto IMBHs
provides a source of (re)ionizing radiation. 
We show that the baryon mass fraction locked into IMBHs and their growth is
strongly constrained by the observed residual Soft X-ray Background (SXRB) 
intensity. Thus, unless they are 
extremely X-ray quiet, miniquasars must be quite rare and/or have a short shining phase. 
As a byproduct, we find that miniquasars can not be the only source of 
reionization and that their alleged contribution to the near infrared bands is
completely negligible.
\end{abstract}


\section{Introduction}

Recent numerical (Bromm, Coppi \& Larson 1999, 2002; Abel, Bryan
\& Norman 2000, 2002) and semi-analytical (Omukai \& Nishi 1998; Omukai 2002;
Nakamura \& Umemura 2001, 2002; Schneider et al. 2002, 2003;
Omukai \& Palla 2003) studies consistently predict that the first,
so-called Population III (PopIII), stars had characteristic masses
of 100-600 $\Msun$, i.e. roughly ten times more massive than those observed
today. Since mass loss at zero metallicity can be neglected (Kudritzki 2002),
the final fate of PopIII stars is essentially set by their initial mass. 
In the narrow range of progenitor masses, $140\;\Msun \le M_\star \le
260\;\Msun$, PopIII stars explode as pair instability supernovae 
(SN$_{\gamma\gamma}$) leaving no remnant and polluting the universe with metals
(Heger \& Woosley 2002). Non-rotating stars with masses $<140\;\Msun$ and 
$>260\;\Msun$ will instead collapse directly into  black holes (BH), without 
metal ejection.
These intermediate-mass BHs (IMBHs) may provide the seed for the supermassive
BHs observed in the center of luminous galaxies (e.g., Volonteri, Haardt \&
Madau 2003). If IMBHs accrete material from the 
surrounding medium, they will shine as ``miniquasars'' and may contribute to
the reionization of the Universe at high redshift (Haiman \& Loeb 1998; 
Venkatesan, Giroux \& Shull 2001; Oh 2001; Ricotti
\& Ostriker 2004a,b; Madau et al. 2004).

In this paper, we discuss the contribution of this population of early
miniquasars to various cosmic backgrounds. We show that available observations
can be used to set strong constraints on the radiative proprieties of these 
sources. In particular the Soft X-ray Background (SXRB) is easily
overproduced unless miniquasars are very rare, or, in case more common, 
if their shining phase is very short.
As a byproduct, we show that, in both cases, their expected contribution
to the Near Infrared Background is negligible.

%This paper is organized as follows. In Sect.~2 we describe the background
%intensity calculation and in Sect.~3 the adopted spectrum of miniquasars.
%Sect.~4 and 5 are dedicated to the study of the soft X-ray and Near Infrared
%background from miniquasars, respectively. 
%Finally, in Sect.~6 we discuss our results.

\section{Cosmic Background Radiation}

The mean specific background intensity
$J(\nu_{0},z_{0})$ at redshift 
$z_{0}$ observed at frequency $\nu_0$, 
produced by a population of sources characterized by a comoving emissivity 
$j_\nu(z)$, can be written as
\begin{equation}\label{eq:I}
J(\nu_{0},z_{0})= \frac{(1+z_0)^3}{4\pi}\int^{\infty}_{z_{0}}
j_\nu(z) \frac{dl}{dz}dz,
\end{equation}
where $\nu=\nu_0(1+z)/(1+z_0)$, and $dl/dz$ is the proper line element. 
In the above expression we have neglected, for reasons that will become clear later on, any 
absorption of photons as they propagate in the expanding Universe. 
The source term $j_\nu$ 
can be written, in general, at any given cosmic time $t$, as a convolution of the light curve 
of the sources $l_{\nu}(t)$ with the source formation rate (per unit comoving volume) $dN/dt$:
\begin{equation}\label{eq:eps}
j_\nu(t)=\int_0^t l_{\nu}(t-t')\frac{dN}{dt'}dt'\simeq \tilde l_{\nu}\tau_S (e^{\tau_{lf}/\tau_S}-1) \frac{dN}{dt}(t).
\end{equation}
where $\tau_S=\epsilon \Msun c^2/f_E L_E\simeq 4.4\times10^8\;\epsilon/f_E$ yr 
is the Salpeter time, $\epsilon$ is the accretion radiative efficiency, and 
$L_E\simeq 1.3\times 10^{38} (M_{BH}/\Msun)$ is the Eddington luminosity.
The second approximated equality holds averaging the light curve 
over a typical source lifetime $\tau_{lf}$, leading to a specific luminosity $\tilde l_{\nu}$. 
and assuming a constant source formation rate over such timescale.
The emissivity is then proportional to the accreted mass 
$\Delta M_{acc} = (e^{\tau_{lf}/\tau_S}-1) M_{BH}$, where $M_{BH}$ is the initial
BH mass. Although the {\it frequency integrated} background depends only on the amount of 
accreted mass, and not on the details of the accretion process, the spectral energy distribution (SED) in a given 
narrow band may depend on the properties of the accreting BHs. For example, for fixed $\Delta M_{acc}$, 
a shorter $\tau_{lf}$ would imply a smaller final BH mass (and a larger number of those BHs), with a 
resulting hotter disk component (see next Section 3).   

Note that, since  $\tilde l_{\nu}$ can be written as a fraction $f_E$ of the Eddington 
luminosity (see next Section), $J(\nu_0,z_0)$ is independent of this parameter.
We compute the source formation rate using the Press-Schechter formalism 
(Press \& Schechter 1974) adopting the minimum dark matter halo mass, 
$M_{min}(z)$, computed by Fuller \& Couchman (2000). 

Throughout the paper, 
we adopt the `concordance' model values for the cosmological parameters:
$h=0.7$, $\Omega_m=0.3$, $\Omega_{\Lambda}=0.7$, $\Omega_b=0.044$,
$\sigma_8=0.9$, and $\Gamma=0.21$, where $h$ is the dimensionless Hubble
constant, $H_0=100h$ km s$^{-1}$ Mpc$^{-1}$

\section{Miniquasar spectrum}

The physical characterization of the sources is encoded in the 
SED $\tilde l_\nu$, and in the typical lifetime $\tau$. The UV/X--ray 
SED of an accreting BH, is, observationally,  
approximatively described in terms of two main continuum components (see, e.g., Tanaka \& Levin 
1995). 
The low energy component is thought to originate from the putative accretion disk, and, at least in 
stellar-sized BHs (the so-called galactic black hole candidates GBHCs), 
it is spectroscopically well described by a ``multicolor disk black body" 
(Mitsuda et al. 1984). The accreting gas is assumed to be optically thick, and 
the gravitational power locally released as black body radiation. Different radii radiate at different 
temperatures, with the hottest Planckian emitted at $\simeq 5 R_S$, where $R_S=2GM/c^2$. Assuming 
Eddington limited accretion, this model yields to $kT_{max}\approx 1\mbox{ keV}
(M_{BH}/\Msun)^{-1/4}$ (Shakura \& Sunyaev 1973). The characteristic multicolor
disk spectrum is broadly peaked at $E_{\rm peak}\approx 3 kT_{max}$,  
follows a power law with $\tilde l_{\nu,MCD}\propto \nu^{1/3}$ for energies 
$h\nu < E_{\rm peak}$, and exponentially rolls off for $E> E_{\rm peak}$. 

The second observed component is a simple power-law 
$l_{\nu,PL}\propto \nu^{-\alpha}$ with $\alpha\simeq 1$. 
The precise origin of this power-law
emission is controversial (several different models exist to date),  
and is likely to originate from synchrotron and/or inverse Compton (IC) emission by a mixture 
of thermal and nonthermal electrons, located in an active corona and/or in a relativistic jet 
(for e recent review see, e.g., Zdziarski \& Gierlinski 2004). Both the low and the high energy ends of the power law 
are not firmly known. Hard--states of GBHCs and Seyfert I galaxies show exponential cut-offs 
at $E\sim 100-500$ keV (Matt 2001), 
while in the steeper soft state of GBHCs, and in more distant objects such as QSOs, observations fail due 
to the low S/N. The low energy end of the power law, as well, could start at $E\sim E_{\rm peak}$, 
if the soft photon input for the IC cooling of the relativistic electrons is due to disk radiation, or, 
instead, could extend into the IR if 
local synchrotron radiation is, in the corona/jet, energetically important (SSC models).   

The two component spectrum, though historically motivated by studies of 
GBHCs and Seyfert galaxies, 
approximates quite well the high energy emission observed in 
ultraluminous X-ray sources, a recently discovered population of X-ray sources possibly associated to 
accreting IMBHs, $M_{\rm BH} \simeq 500-10^4\;\Msun$
(Miller et al. 2003). Observationally, an IMBH seems to have comparable MCD and PL luminosities,  
although the ratio between the two components might vary among sources, and  
the extension of the PL is not known.
In order to model in a simple, but physically motivated, way 
the SED of an accreting IMBH, we assume that the PL component is due to IC scattering 
of thermal disk photons, and hence 
we consider energies $E\ge E_{\rm peak}$. This is a clear difference with respect to the 
SED used by Dijkstra et al. (2004), and by Madau et al. (2004), who assumed the PL component 
to be present for $E\geq 13.6$ eV, so that
it is dominant with respect to the MCD in terms of emitted ionizing
photons. 

We account for the population spread of the PL/MCD flux ratio by introducing an empirical parameter: 
\begin{equation}
\Phi=\frac{L_{PL}}{L_{MCD}}=\frac{\int l_{\nu,PL} d\nu}{\int l_{\nu,MCD} d\nu}. 
\end{equation}
It may worth noticing that, for a fixed BH mass, the value of $\Phi$ determines the 
luminosity of the X--ray part of the emitted spectrum. 
We also parameterize the bolometric luminosity (i.e., MCD+PL) of an IMBH of mass $M_{BH}$ 
as a fraction $f_{E}$ of the Eddington 
luminosity, $L_{E}\simeq 1.3 \times 10^{38} (M_{BH}/\Msun)$ erg s$^{-1}$. 

\section{The Soft X-ray Background}\label{sec:sxrb}

Moretti et al. (2003) determined the intensity of the total SXRB in the
energy range 0.5-2 keV to be $(7.53\pm 0.35)\times 10^{-12}$ erg cm$^{-2}$
s$^{-1}$ deg$^{-2}$ when combining 10 different measurements reported in the
literature. Including further deep pencil beam surveys together with wide
field shallow surveys, they find that $94^{+6}_{-7}$\% of the SXRB is made
up of discrete X-ray sources (the majority being point sources) at low
redshift, $z<4$ (Barger et al. 2002, 2003).  By reanalyzing
the Moretti et al. uncertainty budget, Dijkstra, Haiman \& Loeb (2004) 
provided a mean and a maximum intensities of the unaccounted SXRB flux,  
$0.35\times 10^{-12}$ and $1.23\times10^{-12}$ 
erg cm$^{-2}$ s$^{-1}$ deg$^{-2}$, respectively.

A population of IMBHs forming at high redshift can contribute to the SXRB. 
We compute here the expected background intensity in the energy band 0.5-2 keV
according to Eqs.  (\ref{eq:I})-(\ref{eq:eps}) for a population 
of miniquasars whose formation continues down to a given redshift, $z_{end}$, 
as a final evolutionary product of massive PopIII stars. Given the hard energy band and
low IGM metallicities we are concerned with,  
we neglect any absorption term in the radiative transfer equation (eq.~\ref{eq:I}).  

\begin{figure}
\center{{
\epsfig{figure=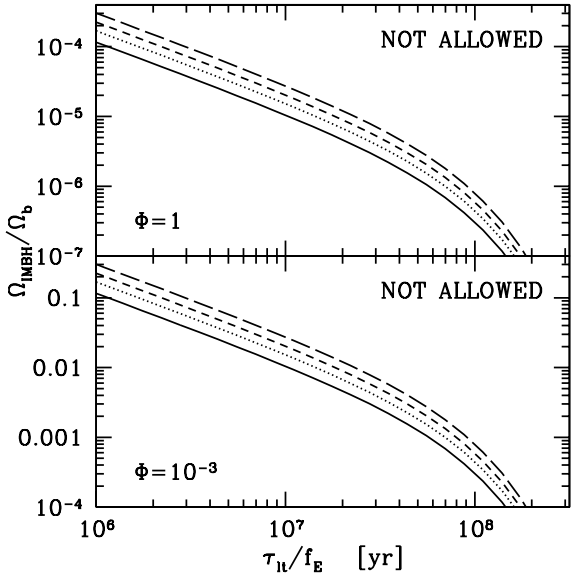,height=8cm} }}
\caption{\label{fig:constraint} Constraints on the IMBH lifetime, $\tau_{lt }$,
 and $\Omega_{IMBH}/\Omega_b$  for redshift $z_{end}=6$ 
({\it solid line}), $z_{end}=12$ ({\it dotted}),  $z_{end}=18$ 
({\it dashed}), and $z_{end}=24$ ({\it long-dashed}). 
The top (bottom) panel refers to the case $\Phi=L_{PL}/L_{MCD}=1$ ($\Phi=10^{-3}$). We adopt here $\epsilon=0.1$.
}
\end{figure}

In the 0.5--2 keV energy range, the
background intensity is dominated by the power-law component of the miniquasar
spectrum, unless $\Phi\ll 1$. We consider here the two extreme values of 
$\Phi=1, 10^{-3}$. 
In order to set upper limits on the propriety of this  first population
of accreting IMBHs, we give the result for the maximum unaccounted SXRB flux 
derived by Dijkstra et al. (2004). 

Unless miniquasars are extremely X-ray quiet (and therefore $\Phi \ll 1$), 
we find that the SXRB sets strong constraints on the density of miniquasars,
which are summarized in the top panel Fig. 1 as a function of their lifetime.
The curves refer to different turn-off redshifts $z_{end}= 6, 12, 18, 24$, respectively.
Apart from the differences in the value of $z_{end}$, which introduces 
an uncertainty of a factor $\approx 2$ on the estimates, we see that  
for $\Phi=1$, the mass fraction of IMBH cannot exceed $10^{-4}\Omega_b$ (that
is of the same order of the density of the SMBH today, Merritt \& Ferrarese
2001), even
for an extremely short lifetime $\tau_{lt} \simeq 10^6$~yr.
This constraint increases to $\Omega_{IMBH}<0.1\;\Omega_b$ in the case in 
which IMBHs are extremely inefficient X-ray emitters ($\Phi=10^{-3}$), i.e.
a large fraction of the baryon density might be locked into BHs without exceeding 
the SXRB constraint (bottom panel of Fig. 1).
%For $\Phi\approx 1$ and $z_{end}=9$, we find $K \sim 1.7\times 10^{-6}$.  
%At higher redshift, the limits on $K$ are weaker, being $\sim 1.4\times 10^{-4}$ 
%($8.4\times 10^{-6}$) for $z_{end}=24$ ($z_{end}=15$).  
Since we have assumed the maximal SXRB residual intensity, the above values
must be understood as strong upper limits. In addition, other sources, such as high
redshift quasars (Dijkstra et al. 2004), may contribute to the unresolved
SXRB, leading to even more stringent limits . 
We conclude that early miniquasars were quite rare and/or their shining 
phase lasted only for a very short period of time.

\begin{figure}
\center{{
\epsfig{figure=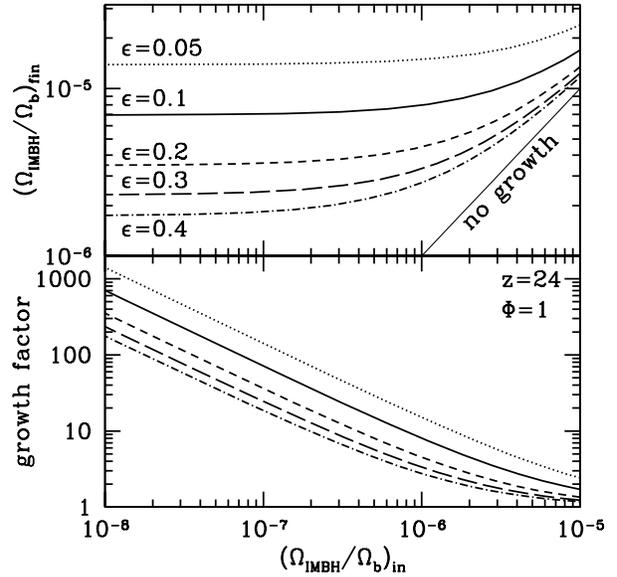,height=8cm} }}
\caption{\label{fig:grow} Maximal growth of IMBHs as function of the initial
mass density of IMBHs forming at $z=24$. 
Different lines refer to different value of $\epsilon$. Top
panel: final density in BHs. Bottom panel: growth factor.}
\end{figure}

\medskip
We can rewrite these limits in terms of the maximal mass growth of BHs allowed
by the unaccounted SXRB. In Fig. \ref{fig:grow} we show the final IMBH density 
and the growth factor, defined as the ratio between final and initial IMBH
density, for sources forming at $z=24$ (different lines
refer to different values of accretion radiative efficiency $\epsilon$) 
as function of the initial mass density of IMBH, $\Omega_{IMBH,in}$. 
For sources forming down to $z_{end}=12$ $(z_{end}=6)$  the limits are 
tighter of a factor 1.8 (2.6).

We find that for $\epsilon=0.1$ a strong upper limit to the final mass density 
$7\times 10^{-6}\;\Omega_b$ for a wide range of 
initial densities, i.e. for $\Omega_{IMBH}/\Omega_b \simlt 10^{-6}\;\Omega_b$. 
Considering that low redshift ($z \simlt 6$) accreting BHs are taken into account
in the resolved fraction of the SXRB, we can derive an upper limit on the
density of BHs (active and not) at $z\simgt 6$ of 
$\rho_{IMBH}<3.8\times 10^4\;\Msun$ Mpc$^{-3}$, which is $\simeq 10$\% of present day SMBH mass density 
(Yu \& Tremaine 2002).
This value is not at odd with
current models of SMBH assembly in a hierarchical structure formation scenario 
(Volonteri et al. 2003; Madau et al. 2004). 
As example, Volonteri et al. (2003) find that the
mass density locked into BHs is of the order of $10^4\;\Msun$ Mpc$^{-3}$ at
$z\sim 6$. This mass growth corresponds to a SXRB contribution of 
$\sim 0.29\times 10^{-12}$ erg cm$^{-2}$ s$^{-1}$ deg$^{-2}$, that is a not 
negligible fraction of the maximal unresolved intensity, and it is comparable to the 
mean value of this quantity. Moreover, we find also a contribution in
the hard-X band (2--10 keV, so--called HXRB) of the order of $0.33\times 10^{-12}$ erg cm$^{-2}$ 
s$^{-1}$ deg$^{-2}$, corresponding to $\simeq 1.7$\% of the observed HXRB.
Thus, within hierarchical clustering models, a significant fraction of
the unaccounted SXRB (and HXRB) should come from the 
growth of IMBHs in the early Universe. For example, using the 
Madau et al. (2004) model we find that the SXRB (HXRB) is 
$\simeq 0.20\, (0.23) \times 10^{-12}$ erg cm$^{-2}$ s$^{-1}$ deg$^{-2}$ 
already at $z=14$ for 3.5$\sigma$ peak seeds,
assuming an accreted mass corresponding to $10^{-3}$ of the halo mass. 

Stronger constraints on the maximal growth of IMBHs in the early Universe can
be set by the next generation of X-ray satellites (e.g. Constellation-X, XEUS)
that will be able to resolve sources 10 times fainter than the present 
facilities. Extrapolating the $\log N/ \log S$ to this flux limit will allow
to resolve the SXRB entirely (Moretti et al. 2003). If this were the case, 
strong limits on the grow history of BHs in the early Universe can be derived, 
or conversely, on the spectral energy distribution at high energies of these
sources. 

\section{Limits on reionization}

The limits on the radiative proprieties of miniquasars derived from the SXRB
constraint allow us to give a simple estimate of the role of these sources in
the reionization of the Universe. 

\begin{figure}
\center{{
\epsfig{figure=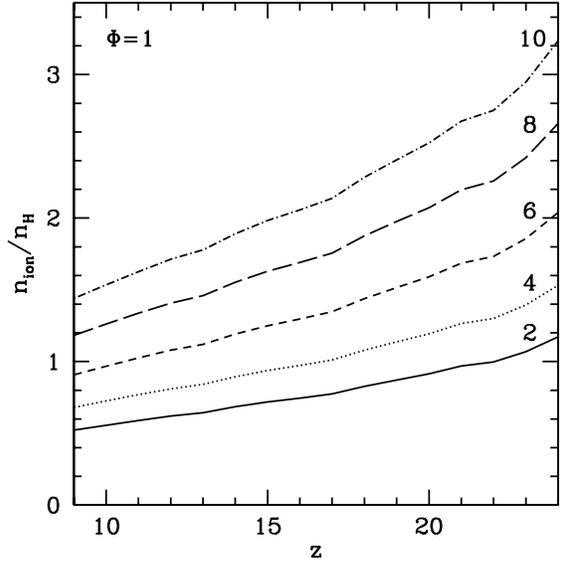,height=8cm} }}
\caption{\label{fig:nion} Limits on the number of ionizing photons per 
hydrogen atom as function of redshift given by the unresolved SXRB.
Different lines refer to different lifetime, $\tau_{lf}$. The labels report the
number of e-folding $\tau_{lf}/\tau_S$. 
We adopt here
$\epsilon=0.1$.}
\end{figure}

The number of ionizing photon per hydrogen atom can be written as 

\begin{equation}
\frac{n_{ion}}{n_H}=\frac{\epsilon m_H c^2}{X} \frac{f_{UV}}{\langle h\nu \rangle}
\left(\frac{\Omega_{IMBH}}{\Omega_b}\right)_{fin}
\end{equation}
where $(\Omega_{IMBH}/\Omega_b)_{fin}=(e^{\tau_{lf}/\tau_S}-1)(\Omega_{IMBH}/\Omega_b)_{in}$,
$f_{UV}$ is the fraction of the bolometric power emitted as 
hydrogen-ionizing photons with mean energy $\langle h \nu \rangle$,
$m_H$ is the hydrogen mass, and $X\simeq 0.76$ is the mass fraction in 
hydrogen. In Fig. \ref{fig:nion} are shown the limits on this quantity derived
from the maximum unaccounted SXRB assuming $\epsilon=0.1$. Different
lines refer to different lifetime of the miniquasar phase, $\tau_{lf}$. 
The labels report the number of e-folding times, $\tau_{lf}/\tau_S$. 
 
Sources forming at very high redshift and accreting for 10 e-folding
time will be able to produce just 3 photons per hydrogen atoms, but shorter
lifetimes give lower $n_{ion}/n_H$, indicating that miniquasars can not
easly reionize the Universe if recombination is taken into account.
At lower redshift the situation is even worst. Miniquasars forming down to 
redshift 9 can not produce more than one ionizing photon per hydrogen atom 
without saturating the SXRB. Moreover, these values are to be taken as
strong upper limits, since we considered the maximum residual SXRB intensity. 
Using the mean unaccounted SXRB intensity will lead to a reduction
of a factor $\sim 1/3$ of $n_{ion}/n_H$.

In conclusion, miniquasars unlikely account for the reionization
of the Universe even at high redshift, unless they are extremely X-ray quiet. 
In order to have $n_{ion}/n_H\sim 10$ without exceeding the maximum 
unresolved SXRB we must required $\Phi<0.15$ (0.07), 
for sources forming down to redshift $z=24$ ($z=9$) and living 
$\tau_{lf}\sim 4 \tau_S$.

\section{The Near Infrared Background}

Recent measures of the Near Infrared Background (NIRB; see Hauser \& Dwek 2001
for a review) show an excess with respect of the observed light from galaxies
in deep field surveys (Madau \& Pozzetti 2000; Totani et al. 2001).
The discrepancy is maximal in the J band corresponding to 
$1.7-4.8 \times 10^{-5}$ erg s$^{-1}$ cm$^{-2}$ sr$^{-1}$. The large 
uncertainty on this value is given by the different
adopted subtraction of zodiacal light (i.e. sunlight scattered by
the interplanetary dust) contribution\footnote{The lower limit
is obtained for the zodiacal light  model of Wright (1998), 
whereas the upper limit is for the Kelsall et al. (1998) one.}. 

Estimates based on theoretical models suggest that this excess can be well
produced by redshifted light from the first very massive ($M>100\;\Msun$) 
PopIII stars (Santos, Bromm \& Kamionkowski 2002; Salvaterra \& Ferrara 2003) 
if these form efficiently down to $z_{end}=9$. The same stars can also account 
for the observed small scale angular fluctuations detected in the same bands 
(Magliocchetti et al. 2003). In order to avoid over-enrichment of the IGM with 
metals at high redshift, most of these massive stars must end up into IMBHs.
Cooray \& Yoshida (2004) have speculated that if these IMBHs accrete matter and shine as 
miniquasars, they might give an important contribution to the NIRB.

We have revisited this conclusion in the light of the results of the Section
\ref{sec:sxrb}. 
Using the limits implied by the SXRB, we find that the contribution to
the NIRB of these sources is completely negligible. In fact, for $z_{end}=9$, 
the NIRB contribution from  miniquasars in the J band allowed by the SXRB 
constrain is  $ \la 10^{-9}$ erg s$^{-1}$ cm$^{-2}$ sr$^{-1}$, 
hence well below the observed value.

An appreciable contribution in the NIR bands is possible only in the unlikely
case in which miniquasars are extremely X-ray quiet. In this case, $\Phi\ll 1$
and the limits set by the SXRB are very weak. For sources forming down to 
$z_{end}=9$ the NIRB excess data can be fitted without 
saturating the unaccounted SXRB. 
%K=0.04
%Per $\epsilon=0.1$ 9 e-folding times (8103) and initial density 7.5d-6 Omega_b
By redshift 9 and considering $\epsilon=0.15$, 
BHs can increase their mass density for almost 6 e-folding 
times, or about a factor 400. In this case,
an initial BH mass density of $10^{-4}\;\Omega_b$ is sufficient to reproduce
the NIRB data. On the other hand these BH have to accrete all the time down
to $z=9$ resulting in a final mass density of $\Omega_{IMBH}\sim 0.07\;\Omega_b$.

\section{Discussion}

We have studied the contribution of the first generation of miniquasar to
cosmic backgrounds. In particular, we have shown that the observed
residual SXRB intensity (Moretti et al. 2003, Dijkstra et al. 2004) 
can be used to set strong constraints on the abundance and radiative efficiency
of these sources. Unless these objects are extremely X-ray quiet, the SXRB 
is easily overproduced, requiring miniquasars to be quite rare and/or have
a short shining phase. Should accreting IMBHs saturate the SXRB, they would 
contribute also 6-7\% of the HXRB.

As a consequence of our analysis, it is unlikely that miniquasar can  
reionize the Universe, since their are limited to produce $\simlt 3$  
photons per hydrogen atom, even at high redshift. This conclusion is similar
to Dijkstra et al. (2004), though our limits are tighter
owing to a more physically motivated description of the miniquasar spectrum. Moreover
our approach allow us to follow the evolution of the mass density of accreting
IMBHs with time, so that we can derive important constraints on the mass
growth of these objects in the early Universe. We derived a strong upper limit
to (active and not) IMBHs mass density at $z\simgt 6$, being 
$\rho_{BH}< 3.8\times 10^4\;\Msun$ Mpc$^{-3}$. Although this value is not
at odd with current model of SMBH assembly in the hierarchical scenario of 
structure formation, stronger constraints on the SXRB unaccounted fraction
by future X-ray facilities (i.e. Constellation-X, XEUS) could question our
ideas of the formation of quasars. In fact, given the prediction of these
models we expect that a not negligible fraction of the SXRB will not be 
resolved, being the signature of the growth of IMBHs in the early Universe.

As a further byproduct, we have shown that their contribution
in the near infrared bands is completely negligible.\footnote{Similar 
conclusions were reached independently in a similar analysis by Madau \& Silk
(2005)}
In the proposed models of the NIRB (Santos et al. 2002; Salvaterra \& Ferrara 
2003, Magliocchetti et al. 2003) the NIRB excess is due to redshifted light of PopIII 
stars with masses larger than 100 $\Msun$. In order to avoid over-enrichment
of the IGM at high redshift, most of these stars must end up in IMBHs, 
locking $\approx 10$\% of the baryons into compact objects already at $z=9$.
Though not excluded by any of the 
current experiments (including gravitational lensing data, Wambsganns 2002),  
this requirement is somewhat extreme, as pointed out by Madau \& Silk (2005). 
On the other hand, we have shown that miniquasars powered by accretion onto IMBH cannot contribute 
appreciably to the NIRB, as they easily exceed the SXRB constraints. In fact,  
the IMBHs left over by the first stars must be characterized by 
a very short shining phase ($< 10^3 f_{E}^{-1}$ yr, assuming $\Phi \approx 1$)
and/or very low accretion efficiency in order not to overproduce the 
SXRB. As a consequence, IMBHs cannot grow appreciably in mass.
Only in the unrealistic case $\Phi \ll 1$, the contribution to the unaccounted NIRB from 
miniquasars might dominate that of the progenitors. In this case, we found 
that $\sim 7\%$ of the baryons must be locked into IMBHs at $z=9$.

\section*{ACKNOWLEDGMENTS}
We thank P. Madau, M. Ricotti \& M. Volonteri for valuable discussions. We thank the 
anonymous referee for the important comments that have improved the quality
of the paper.

\end{document}